\documentclass[twoside]{article}
\usepackage{qic,epsfig}
\usepackage{graphicx}
\usepackage{color} 
\usepackage{amsmath}
\usepackage{amssymb}
\usepackage{bbold}

\newcommand{\bra}[1]{\langle#1|}
\newcommand{\ket}[1]{|#1\rangle}

\newcommand{\ketbrad}[1]{|#1\rangle\!\langle #1|}
\newcommand{\mb}[1]{\mbox{\boldmath$#1$}}

\def\Ket#1{\left|#1\right\rangle} 
\def\Bra#1{\left\langle#1\right|}
\def\KetBra#1#2{\Ket{#1}\!\Bra{#2}} 

\def\Proj#1{\KetBra{#1}{#1}}
  
\def\Eins{\mathbf{1}} 
\def\id{\mathbf{1}} 

\def\ie{i.\,e.\ }

\newcommand{\ea}{\emph{et al.}}

\DeclareMathOperator{\tr}{tr}
\DeclareMathOperator{\vrspan}{span}

\def\example{\noindent\emph{Example:\ }}

\def\ia{a}
\def\ib{b}
\def\H{\ensuremath{\mathcal{H}}} 
\def\B{\ensuremath{\mathcal{B}}} 
\def\V{\ensuremath{\mathcal{V}}} 
\def\parties{\ensuremath{{P}}} 
\def\tensortimes{}

\begin{document}
\setlength{\textheight}{8.0truein}    

\runninghead{Local invariants for multi-partite entangled states  $\ldots$}
            {H. Aschauer, J. Calsamiglia, M. Hein, H. J. Briegel}

\normalsize\textlineskip
\thispagestyle{empty}
\setcounter{page}{1}

\copyrightheading{0}{0}{2003}{000--000}

\vspace*{0.88truein}

\alphfootnote

\fpage{1}

\centerline{\bf
Local invariants for multi-partite entangled states}
\vspace*{0.035truein}
\centerline{\bf allowing for a simple entanglement criterion}
\vspace*{0.37truein}
\centerline{\footnotesize Hans Aschauer$^{1}$, John Calsamiglia$^{2}$, 
Marc Hein$^{1,2}$, and Hans J. Briegel$^{1,2,3}$}
\vspace*{0.015truein}
\centerline{\footnotesize\it $^1$Sektion Physik, 
Ludwig-Maximilians-Universit\"at M\"unchen}
\centerline{\footnotesize\it Theresienstr.\ 37, D-80333 M\"unchen, Germany}
\baselineskip=10pt
\centerline{\footnotesize\it $^2$Institut f\"ur Theoretische Physik,
Universit\"at Innsbruck,}
\centerline{\footnotesize\it A-6020 Innsbruck, Austria.}
\baselineskip=10pt
\centerline{\footnotesize\it $^3$Institut f\"ur Quantenoptik und
Quanteninformation der
\"Osterreichischen Akademie der Wissenschaften}
\centerline{\footnotesize\it A-6020 Innsbruck, Austria.}
\vspace*{0.225truein}
\publisher{(received date)}{(revised date)}

\vspace*{0.21truein}

\abstracts{
We present local invariants of multi-partite pure or mixed states,
which can be easily calculated and have a straight-forward physical
meaning. As an application, we derive a new entanglement criterion for
arbitrary mixed states of $n$ parties. The new criterion is weaker than the 
partial transposition criterion but offers  advantages for the 
study of multipartite systems. A straightforward generalization 
of these invariants allows for the construction of a
complete set of observable polynomial invariants.
}{}{}

\vspace*{10pt}

\vspace*{3pt}
\communicate{}

\vspace*{1pt}\textlineskip    


\section{Introduction}
\label{sec:Intro}

Quantum mechanical states have a complex description in terms of their
density matrix, which comprises all information available about a
system under a given experimental situation. Different density
matrices correspond to different states of a system, and allow for
different predictions on its future behavior. For many purposes,
however, we are only interested in properties of the state (such as
its entropy or purity) which are invariant under  unitary
transformations.

For systems composed of several parts, or subsystems, there exists a
natural tensor product structure underlying the state space. For such
composite systems, the superposition principle gives rise to the
phenomenon of entanglement which manifests itself in peculiar
``quantum'' correlations between results of measurements on its
different parts \cite{EPR:35,schroedinger:35,bell:64}.  To capture the
essential features of this entanglement, we look for properties of the
density matrix that are invariant under \emph{local unitary
  transformations}, corresponding to a local change of basis in the
Hilbert spaces of the individual subsystems. Such local invariants have
attracted the attention of people working on the foundation of quantum
mechanics and, more recently, in quantum information theory
\cite{schlienz-mahler:1995:1996,linden-popescu:98,grassl-et-al:1998,
verstraete-et-al:2002,jaeger-et-al:2003},
where entanglement is perceived as a resource for tasks in quantum
communication and computation.

In this paper, we present a family of local invariants of a
multi-partite quantum system. These invariants are derived from an
invariant decomposition of the state space of the system, regarded as
a real vector space of hermitian operators with a scalar product.
They have a natural geometric interpretation in terms of the length of
projections of vectors onto invariant subspaces. Such subspaces
contain all
information either about one local subsystem \emph{or} about
correlations between a given set of subsystems.  Beyond their
straight-forward 
geometric interpretation, these invariants have a number of merits.
They can easily be calculated -- even analytically -- for many states,
and they are directly connected to measurement data
\cite{james-kwiat-et-al:2001,thew-nemoto-et-al:2002}.

The representation of the density matrix as an element in the real
(metric) vector space of hermitian matrices is well known, and a
number of researchers have used a similar approach before
\cite{schwinger:1960,hioe-eberly:1981,schlienz-mahler:1995:1996,englert-metwally:99,zukowski-brukner:2002,Byrd-khaneja:2003}.
Nevertheless, our results go beyond existing work in at least two
respects. First, the explicit decomposition of the state space into a
direct sum of invariant sub-spaces makes the identification of
invariants quite transparent --- it allowed us in fact to find a family
of new invariants.  Second, from the \emph{convexity} of the set of
separable states, we are able to derive constraints on the invariants
of separable states and propose a new entanglement
criterion. Here, we will discuss some of its strengths and limitations 
and apply it to a wide class of multi-qubit states.

Moreover, for each state the obtained invariants are homogeneous
polynomials of degree $2$ in the coefficients of a basis
decompositions of the density operator into hermitian operators. We
show how considering polynomials of higher degree
one can extend the set of local invariants and make it complete
\cite{grassl-et-al:1998}.

\section{State tomography}

It is a well known fact that the four Pauli spin matrices \(\sigma_0 =
\Eins, \sigma_1 = \sigma_x, \sigma_2 = \sigma_y, \sigma_3 = \sigma_z\)
form a real basis of the vector space of the hermitian operators which
act on one qubit. With respect to the scalar product \(\left< A, B
\right> = \tr(A B)\), the basis vectors are orthogonal. More
generally, for a $d$-dimensional quantum system, there exists a set of
$d^2 - 1$ traceless hermitian generators of the $SU(d)$, which we call
\(\sigma_1, \ldots \sigma_{d^2-1}\). One specific choice of these
generators is the so-called Cartan-Weyl-construction (see,
e.\,g. \cite{schlienz-mahler:1995:1996}). Combined with the unit operator
$\Eins \equiv \sigma_0$, they form a real non-normalized orthogonal
basis of the vector space of hermitian operators in $d$ dimensions,
\begin{equation}
  \label{eq:normalization}
  \left\langle\sigma_i, \sigma_j \right\rangle = \tr (\sigma_i
  \sigma_j) =  \delta_{i,j} d
\end{equation}

Let \(P=\{1,2,\ldots,n\}\) be a set of parties and \V\ the vector space of
hermitian operators acting on the n-partite Hilbert space \(\H^{(1)}
\otimes \H^{(2)} \otimes \cdots \otimes \H^{(n)}\), where \(\H^{(a)}\)
is a Hilbert space of the (finite) dimension \(d_a\). Clearly, the
tensor products of the basis operators form a basis
\begin{equation}
  \label{eq:basis}
  \B = \{\sigma_{i_1}^{(1)} \tensortimes \sigma_{i_2}^{(2)} %
    \tensortimes \cdots \tensortimes \sigma_{i_n}^{(n)} | 0\le i_a \le
    d_a^2 - 1 \mathrm{\ for\ all\ } a\in P\}
\end{equation}
of \V.

Any $n$-partite density operator \(\rho \in \V\) can thus be expanded
in the product basis%
\begin{equation}
  \label{eq:expansion_of_rho}
  \rho = \frac{1}{d} \sum_{i_1,i_2,\ldots ,i_n}
    c_{i_1 i_2 \ldots i_n}%
    \sigma_{i_1}^{(1)} \tensortimes \sigma_{i_2}^{(2)} %
    \tensortimes \cdots \tensortimes \sigma_{i_n}^{(n)},
\end{equation}
where \(d = \prod_{a=1}^n d_a\), and
\begin{equation}
  \label{eq:expansion_coefficients}
  c_{i_1 i_2 \ldots i_n} = \tr\left(
    \rho \,\sigma_{i_1}^{(1)} \tensortimes \sigma_{i_2}^{(2)}
    \tensortimes \cdots \tensortimes \sigma_{i_n}^{(n)}\right) =
  \left\langle \sigma_{i_1}^{(1)} \tensortimes \sigma_{i_2}^{(2)}
    \tensortimes
    \cdots \tensortimes \sigma_{i_n}^{(n)} \right\rangle_\rho.
\end{equation}
In other words, the expansion coefficients \(c_{i_1i_2\ldots i_n}\)
are expectation values of products of hermitian operators. Since these
expectation values can, in principle, be measured by local
measurements (given a sufficiently large ensemble of copies of
$\rho$), one can use this method in order to determine an unknown
$n$-partite quantum state with the help of local measurements and
classical communication (quantum state tomography).

\section{Invariant decomposition of the state space}

Let \(\sigma = \sigma_{i_1}^{(1)} \tensortimes \sigma_{i_2}^{(2)} %
\tensortimes \cdots \tensortimes \sigma_{i_n}^{(n)}\) be an arbitrary
element of the product basis \B, and \(S = \{\ia | i_\ia \not=
0\}\) the set of parties, where \(\sigma\) acts non-trivially. Using
this definition, we call \(\sigma\) a \(S\)-correlation operator, and
the set of all $S$-correlation operators \(\B_S\). It is clear that
\B\ can be written as the union of the (disjoint) sets of
$S$-correlation operators, \ie
\begin{equation}
  \label{eq:partitioning}
  \B = \bigcup_{S \subset \parties} \B_S.
\end{equation}

\example In the case of three qubits, we have eight such sets
(with \(\ia, \ib \in \{1,2,3\}\)):
\(\B_{\{\}}= \{\Eins\}\), \(\B_{\{\ia\}}=
\{\sigma_i^{(\ia)}|i=1,2,3\}\), \(\B_{\{\ia, \ib\}}=
\{\sigma_i^{(\ia)}\sigma_j^{(\ib)}|i,j=1,2,3\}\), and
\(\B_{\{1,2,3\}}= \{\sigma_i^{(1)}
\sigma_j^{(2)} \sigma_k^{(3)}  |i,j,k=1,2,3\}\).

\begin{theorem}
  \label{theorem:one}
  For each set $S$ of parties, the vector space \(\V_S =
  \vrspan(\B_S)\) is invariant under local unitary transformations,
  which act as  isometries  on \(\V_S\).
\end{theorem}
\proof{
  Let $U^{(\ia)}$ be a unitary operation which acts on party $\ia$. If
  $\ia\not\in S$, all elements of \(\B_S\) remain unchanged under the
  action of $U^{(\ia)}$. If, on the other hand, $\ia \in S$, then the
  orthogonal set of traceless generators \(\sigma_i^{(a)}\) ($i>0$) is
  transformed into a different set of orthogonal traceless generators,
  \ie for \( 1 \le i \le d_a^2 -1\),
  \[\sigma_i^{(\ia)} \rightarrow \tilde\sigma_i^{(\ia)} =  \sum_k \left(
    O(U^{(\ia)})\right)_{ik} \sigma^{(a)}_k \] with an orthogonal
  matrix \(O(U^{(\ia)}) \in SO(d_a^2-1)\)
  \cite{schwinger:1960,schlienz-mahler:1995:1996}. 
  Obviously, both
  sets of generators span the same set of \emph{all} traceless
  operators.
}

\medskip

Given a density operator \(\rho\) and a set $S$ of parties, the
projection of \(\rho\) onto the subspace \(\V_S\) is
given by
\begin{equation}
  \label{eq:decompose_rho}
  \xi_S(\rho)= \frac{1}{d} \sum_{\sigma\in\B_S} %
      \left\langle \rho, \sigma \right\rangle \sigma  = %
      \frac{1}{d} \sum_{\sigma\in\B_S} %
      \left\langle %
        \sigma
      \right\rangle_\rho\sigma.
\end{equation}
For simplicity, we will often denote the above projection
by $\xi_{S}$, where the dependence on the state $\rho$
in question will be understood.

The reduced density operator $\rho_{S}$ corresponding to the set of 
parties $S$ can be read off,
\begin{equation}
    \label{eq:partial_trace}
  \tilde{\rho}_{S}\equiv \rho_{S}\otimes\frac{\id_{\bar{S}}}{d_{\bar{S}}} = 
    \sum_{S' \subset S}
    \xi_{S'}(\rho)\mbox{,}
\end{equation}
where $\bar{S}=P\backslash S$ denotes the complement of set $S$.

By direct substitution into the previous equation one readily checks
that the following relation between the projection of $\rho$ onto $\V_S$
and its reductions holds
\begin{equation}
    \label{eq:xiS}
\xi_S(\rho)= \sum_{S' \subset S} (-1)^{|S|-|S'|} \tilde{\rho}_{S'}.
\end{equation} 

Due to Theorem~1, local unitary operations rotate a projection
\(\xi_S\) only within the subspace \(\V_S\). Ignoring
the normalization constant \(1/d\) leads us to
\begin{corollary}
  \label{correlation_strength}
  For each set \(S\) of parties, the squared length of the projection
  of a state $\rho$ onto the span $\V_S$ of $\B_S$,
  \begin{equation}
    \label{eq:sq_length}
    L_S(\rho) = \sum_{\sigma\in\B_S} %
    \left\langle %
      \sigma
    \right\rangle_\rho^2=d \tr (\xi_{S}(\rho)^{2})
  \end{equation}
  is invariant under local unitary transformations. We call
  \(L_S(\rho)\) the $S$-correlation strength of \(\rho\).
\end{corollary}

From Eq.(\ref{eq:xiS}) we find that S-correlation strength depends 
solely on the purities of the reductions of $\rho_{S}$,
\begin{eqnarray}
  \label{eq:strength_purity}
  L_S(\rho) &  = & d \tr(\xi_{S}\xi_{S})= d \tr(\xi_{S}\rho) = d \sum_{S'\subset S} (-1)^{|S|-|S'|} 
  \tr(\rho_{S'}\otimes\frac{\id_{\bar{S}}}{d_{\bar{S}}}\rho) \nonumber \\
 &  = &   \sum_{S'\subset S} (-1)^{|S|-|S'|} \frac{d}{d_{\bar{S}}}
  \tr\left[\rho_{S'} \tr_{\bar{S}}\left(\rho\right)\right] =
 \sum_{S'\subset S} (-1)^{|S|-|S'|} d_{S'}
  \tr(\rho_{S'}^{2}) \; ,
\end{eqnarray}
where $d_{S'}=\prod_{\ia  \in S'} d_\ia$ and we have made use of 
$\tr(\xi_{S}\xi_{S'})=d_{S}\delta_{S ,S'}$.
We notice that the S-correlation depends only on the reduced density 
matrix of the set of parties $S$, $L_S(\rho)=L_{S}(\rho_{S})$.
Thus, the only invariant which contains information about the total
state is \(L_\parties(\rho)\).

For pure product states (i.e. $\tr(\rho_{S}^{2})=1$  $\forall S$) we find
\begin{equation}
  \label{eq:strength_pure}
  L_S^\mathrm{pure} = \sum_{S'\subset S} (-1)^{|S|-|S'|} \prod_{\ia
    \in S'} d_\ia= \prod_{\ia\in S} (d_\ia-1),
\end{equation}
which  further simplifies to 
$L_{S}^\mathrm{pure}=1$  for $n$-qubit systems.

The idea of local-invariant spaces and correlation strength can be 
easily extended to a scenario where parties can form coalitions,
giving rise to different \emph{partitions} of the set of parties. 
We allow the parties \(\ia_1\ldots \ia_k\) in a coalition to
apply joint operations, or equivalently, we treat them as a single super-party 
acting on higher-dimensional quantum system $\ib$.
One can obtain the required traceless generators for the new party
\(\ib\) as products of the generators of the old parties \(\ia_1\ldots
\ia_k\),
\begin{equation}
  \label{eq:products_partitions}
  \tilde  \sigma_{i_1\ldots i_k}^{(\ib)} =
  \sigma_{i_1}^{(\ia_1)} 
  \sigma_{i_2}^{(\ia_2)} \cdots  \sigma_{i_k}^{(\ia_k)},
\end{equation}
with \((i_1,\ldots, i_k) \not= (0,\ldots, 0)\).

Any partitioning can be realized by iteratively joining parties
pairwise, say \(\ia_1, \ia_2 \rightarrow \ib=\{\ia_{1},\ia_{2}\}\). Using
Eq.~(\ref{eq:products_partitions}), one can easily verify that the
correlation strength for a set \(S = \{\ib\} \cup S' = \{\ib\} \cup
\{\ia_\mu, \ia_\nu, \ldots\}\) of parties, is given by
\begin{equation}
  \label{eq:partition_strength}
  L_{\{\ib\} \cup S'} = %
  L_{\{\ia_1\}\cup S'} + 
  L_{\{\ia_2\}\cup S'} + 
  L_{\{\ia_1, \ia_2\} \cup S'},
\end{equation}
which means that the correlation strengths for coarse
partitions are functions of the correlations strengths of the finest
partition. 

A special partition is obtained if we allow \emph{all} parties to
operate jointly as a super-party $b=P=\{\ia_{1},\ldots,\ia_{n}\}$.
\(L_{\{\ib\}}\) is then invariant under
\emph{all} unitary operations, and thus describes a global
property of the state.  Indeed, we have
\begin{equation}
  \label{eq:trace_rho_squared}
  L_{\{P\}}(\rho) = \sum_{\sigma\in\B} %
    \left\langle %
      \sigma
    \right\rangle_\rho^2 - \langle \Eins \rangle_\rho %
    = d \tr\left( \rho^2 \right) - 1,
\end{equation}
so that \(L_{\{P\}}\) is a measure for the purity of the state
\(\rho\). 

Using Eq.~\ref{eq:partitioning} and \ref{eq:sq_length}, we can re-write
the left-hand side of Eq.~\ref{eq:trace_rho_squared} as the sum of all
$S$-correlation strengths,
\begin{equation}
  \label{eq:correlation_sum}
  \sum_{\{\} \not= S \subset \parties} L_S = d \tr\left( \rho^2
  \right) - 1,
\end{equation}
which allows us to state 
\begin{corollary}
  \label{cor:trade_off}
  For any state $\rho$, the sum of all correlation strengths is
  given by the purity of the state.
  This implies, in particular, that for states with the same
  purity, there is a trade-off between local and the different
  non-local correlations.
\end{corollary}
For a pure state \(\rho = \Proj{\psi}\), we have \(\tr(\rho^2) =
\tr(\rho) = 1\), so that Corollary \ref{cor:trade_off} can be regarded
as a quantitative expression of the folklore saying that in entangled
states, the information about the state is contained in its
correlations rather than its local properties.

\medskip 
It is a useful fact that the convex structure of the space
$\V$ of states is obeyed by the subspaces \(\V_S\) separately, in the
following sense: If a state is given by a convex sum of states
\(\rho_l\), \ie \(\rho = \sum_l p_l \rho_l\) with \(p_l>0\) for all
$l$ and \(\sum_l p_l = 1\), then the projection of \(\rho\) onto each
of the subspaces \(\V_S\) is the convex sum of the projections of the
states \(\rho_l\) onto \(\V_S\).  If \(\rho\) is a separable state, it
can be written as a convex sum of pure product states. In this case,
the projection of \(\rho\) onto each of the subspaces \(\V_S\) is a
convex sum of vectors with the squared length \(L_S^\mathrm{pure}\),
so that the squared length \(L_S(\rho)\) cannot exceed
\(L_S^\mathrm{pure}\).  This allows us to formulate the following entanglement
criterion:
\begin{theorem}
  \label{corr:entanglement_criterion}
 Given a multi-partite state $\rho$, if  there exists a subset \(S\) of parties such
  that the $S$-correlation strength is greater than
  \(L_S^\mathrm{pure}\), then \(\rho\) is entangled.
\end{theorem}

Corollary~ 2 then implies that all pure multi-partite entanglement 
will be detected by this criterion.

It is interesting to note that the strongest entanglement criterion 
is obtained for the finest partition, in the following sense:
Let \(\ib=\{\ia_1, \ia_2\}\) be a coarsening as in
Eq.~(\ref{eq:partition_strength}), and \(L_S(\rho)
< L_S^\mathrm{pure}\) for all \(S \subset \{\ib_1,\ib_2\} \cup S' \subset
\parties\).  Using (\ref{eq:partition_strength}) for the state
\(\rho\) and for product
states, we find %
\begin{equation}
  \label{eq:strongest_criterion}
  \begin{split}
    L_{\{\ib\} \cup S'} &= %
    L_{\{\ia_1\} \cup S'} + %
    L_{\{\ia_2\} \cup S'} + %
    L_{\{\ia_1, \ia_2\} \cup S'} \\%
    &\le %
    L^\mathrm{pure}_{\{\ia_1\} \cup S'} + %
    L^\mathrm{pure}_{\{\ia_2\} \cup S'} + %
    L^\mathrm{pure}_{\{\ia_1, \ia_2\} \cup S'} \\%
    & =   L^\mathrm{pure}_{\{\ib\} \cup S'}.
  \end{split}
\end{equation}
This means that we do not detect entanglement in any coarse partition,
if we do not detect it  in the finest partition.

The correlation-strength $L_S$ defined here can be understood as special
case of `$\mathtt{g}$-purity' defined by Barnum \ea \cite{barnum04} (see also 
\cite{klyachko02}) as the purity relative to a restricted subset of observables 
$\mathtt{g}$. They use this definition to give a `generalized notion of
entanglement': given a subset of observables $\mathtt{g}$ 
(not necessarily related to a sub-system), a pure state is `unentangled' iff 
the $\mathtt{g}$-purity is maximal. 
Other than using the standard convex roof extension of 
such criterion \textemdash for which no closed form is known\textemdash 
, their criterion is limited to pure states.

The correlation strength 
criterion is also stronger than the criterion that results from the `global'
entanglement measure proposed by Meyer and Wallach\cite{meyer02}.
The latter has in fact  been proven \cite{brennen03} to be equivalent to 
the criterion  proposed in \cite{barnum04} applied to multi-partite systems.

For multi-qubit systems, the generalized 
Bell-inequalities \cite{werner01,zukowski-brukner:2002} \textemdash 
criterion for the existence of a local-realistic description of
a set of correlation measurements on a given state\textemdash 
provides a weaker entanglement criterion than the correlation strengths.
Indeed, if there exists a set of local dichotomic measurements on a 
state $\rho$ such that a Bell-inequality is violated, than it follows 
 that there is a local coordinate system
$\{x,y\}$ such that \mbox{$\sum_{i_{1},\ldots i_{N}=1}^{2} c_{i_{1}\ldots 
    i_{N}}^{2}>1$} (Eq. (15) in \cite{zukowski-brukner:2002}). 
Since this is precisely the sum in Eq.(\ref{eq:sq_length}) excluding 
the terms with $i_{k}=3$, one arrives at $L_{P}(\rho)>1$.

However, as one might have guessed by the fact that our criterion only 
depends on the mixedness of $\rho$ and its reductions, 
the criterion fails to detect  a variety of mixed entangled states.
This will be seen in the next section where some important 
classes of multi-qubit states will be studied.
In fact, it turns out \cite{calsamiglia04} that the proposed criterion
is weaker than the well known positive  partial transposition
(PPT) bipartite separability criteria \cite{peres:96}: 
\ie any state for which a given 
correlation strength $L_{S}(\rho)>1$, will have a bipartite split 
$\{A,\bar{A}\}$ such that the corresponding partial transposition 
results in a non-positive operator $\rho^{T_{A}}\ngeq0 $.

\section{Correlation strengths for multi-qubit systems}

In this section we will compute the local invariants and check 
the entanglement criterion for different classes of relevant $n$-qubit states.
We will see that for most examples the correlation strengths can be 
computed even for large values of $n$.

\subsection{Dicke-States}

Dicke states $\ket{n,m}$ are n-qubit symmetric pure states 
with $m$ excitations (or qubits in state $\ket{0}$),
\begin{equation}
\ket{n,m}\propto\sum_{i}\Pi_{i}\ket{1,\stackrel{m}{\ldots},1,0,\ldots 0}
\end{equation}
with $\{\Pi_{i}\}$ being the group of all permutation
matrices. We denote $\rho^{m}=\ketbrad{n,m}$.
Obviously, the states $\ket{n,m}$ and $\ket{n,n-m}$
will have the same entanglement properties.

Dicke states appear naturally in 
quantum optical and condensed matter systems, and have also received attention within the 
field of quantum information (see for example 
\cite{stockton03},\cite{dur00}).
Due to the large symmetry of these states, the purity of their reduced 
density matrices can be readily computed in 
the excitation-basis leading to,
\begin{equation}
L_{S}(\rho^{m})=\sum_{k=0}^{|S|}2^{k}{|S|\choose k}\sum_{n=0}^{\min 
(k,m)}\left({n \choose m}^{-1}{k\choose n}{n-k \choose m-n}\right)^{2}
\end{equation}
which depends only on the size $|S|$ of the set of parties.

For $m=1$, which corresponds to $W$ states \cite{dur00}, we obtain a 
S-correlation strength
 $L_{S}(\rho^{m=1})=\frac{1}{n^{2}}(n^{2}+8 |S|^{2}-4(n+1)|S|)$.
 The maximum value is achieved for the correlation strength involving 
 all parties $L_{P}(\rho^{m=1})=5-\frac{4}{n}$.
 Entanglement in such states is know to be extremely robust to particle
 loss, exhibiting entanglement even when all but two particles are lost.
 However, we see that using our criterion,
 entanglement is only detected for sets $S$ with more than 
 $|S|>\frac{n+1}{2}$ parties.
 Similarly, for other $m$ values we obtain maximum values that converge 
 as $\frac{1}{n}$ for increasing $n$ and also exhibit a threshold for 
 the minimal size of the sets of parties to detect entanglement.
 

\subsection{Graph and Graph-diagonal states}
Graph-states \cite{hein03,graph-codes}  form a new 
and very wide class of multi-partite entangled states 
that have been shown to play an important role in
quantum information processing: they occur in some error correcting
 codes \cite{graph-codes}, 
provide a universal resource for quantum computation \cite{raussendorf01},
 improve frequency standards \cite{huelga97}, and allow for secret sharing  
 \cite{secretsharing}.
Something particularly interesting in the context of studying multi-partite 
entanglement is that they count with an efficient description.
Given a graph $G=(V,E)$ \textemdash \ie a set of $n$ vertices $V$
connected by edges $E$ that specify the neighborhood relation between 
vertices\textemdash graph-states can be conveniently defined through the set of commuting 
observables $\{K_{G}^{a}\}_{a=1}^{n}$,
\begin{equation}
 K_{G}^{a}=\sigma_{x}^{(a)}\prod_{b\in 
    N_{a}}\sigma_{z}^{(b)} \; \mbox{ for } a=1,\ldots,n
\end{equation}
where $N_{a}$ denotes the set of neighboring 
vertices of vertex $a$ ($\{b:\{a,b\}\in E\}$).
The common eigenvectors of this set of observables form a complete orthonormal basis of 
graph-states corresponding to the graph $G$,
\begin{equation}
K_{G}^{i}\ket{\psi_{\vec{\mu}}}=(-1)^{\mu_{i}}\ket{\psi_{\vec{\mu}}} 
\mbox{ for  } i=1,\ldots, n
\end{equation}
where the $n$-dimensional binary array $\vec{\mu}$ labels each 
of the $2^{n}$ graph basis states.

These basis vectors are related by a local unitary 
$\ket{\psi_{\vec{\mu}}}=\sigma_{z}^{\vec{\mu}}\ket{\psi_{\vec{0}}}$, 
where $\sigma_{z}^{\vec{\mu}}$ denotes the action of $\sigma_{z}$ 
on the parties $i$ specified by $\mu_{i}=1$.
 
The stabilizer $S_{G}$ of a graph-state is the finite group generated 
by $\{K_{G}^{a}\}_{a=1}^{n}$. Every element in $S_{G}$ will be a Pauli 
operator that can be labeled by a binary array $\vec{\nu}$:
$\sigma_{\vec{\nu}}=\prod_{a=1}^{n}(K_{G}^{(a)})^{\nu_{a}}$.

Thus, in the Pauli operator basis the representative graph state is 
given by $\ketbrad{\psi_{\vec{0}}}=\frac{1}{d}\sum_{\sigma_{\vec{\nu}}\in 
S_{G}}\sigma_{\vec{\nu}}$, and the other basis states by 
$\ketbrad{\psi_{\vec{\mu}}}=\frac{1}{d}\sum_{\sigma_{\vec{\nu}}\in 
S_{G}}(-1)^{\vec{\mu}\cdot\vec{\nu}}\sigma_{\vec{\nu}} $.
Thus, we notice that the $S$-correlation strength of a graph states
 is given by the number of elements of the 
stabilizer that are $S$-correlation operators, \ie
$L_{S}(\ketbrad{G})=|S_{G}\cap\B_{S}|$. 

By making use of Eq. (\ref{eq:strength_purity})
and the fact that graph states have equally weighted Schmidt 
coefficients for any bi-partition \cite{hein03} we can also rewrite 
the $S$-correlation length as,
\begin{equation}
L_{S}(\ketbrad{G})= \sum_{S'\subset S} (-1)^{|S|-|S'|} 
2^{|S'|-E^{S'}(G)} 
\end{equation}
where the Schmidt measure $E^{S'}(G)$ in respect to partition
$(S',\bar{S'})$ is determined by the Schmidt rank $r$, 
$E^{S'}(G)=\log_{2}(r)$ \cite{hein03}.

An arbitrary n-qubit state can be depolarized by 
local operations and classical communication to a graph-diagonal 
\cite{duer-aschauer-briegel:2003}
state. Specifically, this is done by applying sequentially the mixing maps 
$\mathcal{D}_{i}(\rho)=\frac{1}{2}(\rho+ K_{G}^{i} \rho K_{G}^{i})$ for 
$i=1,\ldots,n$.
For such a graph-diagonal state $\rho_{G}$ with weights $\{p_{\vec{\mu}}\}$ 
the $S$-correlation strength is given by
\begin{equation}
    L_{S}(\rho_{G})=\sum_{\sigma_{\vec{\nu}}\in 
    S_{G}\cap\B_{S}}\left(\sum_{\vec{\mu}}(-1)^{\vec{\nu}
    \cdot\vec{\mu}}p_{\vec{\mu}}\right)^{2}=\mb{p}\cdot M\cdot \mb{p}
    \label{eq:LSG}
\end{equation}
with $M_{ij}=\sum_{\sigma_{\vec{\nu}}\in 
S_{G}\cap\B_{S}}(-1)^{\vec{\nu}\cdot (\vec{S}_{i}+\vec{S}_{j})}$

This can be readily computed numerically and in some cases also
analytically. Below we discuss the class of GHZ-diagonal states.

\textbf{GHZ-diagonal states}: 
The generalized GHZ-states \cite{ghz:89} correspond to 
`star'-graphs \cite{hein03}. The representative graph state is given by,
$\ket{\psi_{\vec{0}}}=\frac{1}{\sqrt{2}}(\ket{0}_{z}\ket{0\ldots 
   0}_{x}+\ket{1}_{z}\ket{1\ldots 1}_{x})$.
The remaining GHZ-basis states are $\ket{\psi_{\vec{\mu}}}$, 
where the first bit of the binary array $\vec{\mu}$ 
determines the phase, while the remaining $n-1$ bits correspond to the bit-flips in the $n-1$ last qubits.
Here, we will label each graph-state basis vector either by the 
the binary array $\vec{\mu}$ or by the corresponding integer in the 
decimal basis $\mu\in [0,2^{n}-1]$.

Since for GHZ-states the Schmidt measure is $E^{S}(\ket{GHZ})=1$ for all 
partitions with exception of $S=\{ \}$ and $S=P$ for which it 
vanishes, one can readily show that (Eq.\ref{eq:LSG})
\begin{equation}
    L_{P}(GHZ)=2^{n-1}+\delta_{n,\mathrm{even}} \; \mbox{ and }\;
    L_{S}(GHZ)=\delta_{|S|,\mathrm{even}} \mbox{ for } S\subsetneq P.
\end{equation}

Following the previous notation for the stabilizer group elements, the 
set of Pauli operators $\sigma_{\vec{\nu}}\in S_{G}\cap\B_{P}$ is given 
by 
$\mathcal{A}^{\mathrm{odd}}=\{\vec{\nu}\}=\{(1,\vec{x})|x=0,\ldots ,
2^{n-1}-1\}$ for 
$n$ odd, and $\mathcal{A}^{\mathrm{even}}=\mathcal{A^{\mathrm{odd}}}\cup \{(0,1,\ldots,1)\}$.
The $P$-correlation strength for GHZ-diagonal states can then be shown 
to be
(Eq.\ref{eq:LSG}),
\begin{equation}
    L_{P}(\rho)=2^{n-1}\sum_{\mu=0}^{2^{n-1}-1}(p_{\mu}-p_{\mu+2^{n-1}})^{2}  
    +\delta_{n,\mathrm{even}} 
    \left(\sum_{\mu=0}^{2^{n-1}-1}(p_{\mu}+p_{\mu+2^{n-1}})(-1)^{|\vec{\mu}|}
    \right)^{2}\mbox{,}\label{eq:LSGGHZ}
\end{equation}
where we have used $\sum_{\vec{x}}(-1)^{\vec{x}\cdot\vec{\mu}}=
\delta_{\vec{\mu},\vec{0}} 2^{n}$.
According to our chosen convention, the state corresponding to the index-label
$\mu+2^{n-1}$ is the same than that with 
label $\mu$ but with opposite phase ($\pm$). Hence, the weights 
$p_{\mu}$ and $p_{\mu+2^{n-1}}$ of GHZ states with opposite phase always appear
together as a difference (and also as sum in the even $n$ case).
Since the reductions of a GHZ are always separable, it is clear that
$L_{S}(\rho)\leq 1$ for $S\subsetneq P$.

These expressions can be further simplified if we restrict to a still 
very relevant class that  can be obtained from a general state 
by a further local depolarization 
step \cite{dur00b}. The class is fully specified by $2^{n-1}$ parameters,
\begin{equation}
    \label{eq:piW}
p_{i}=p_{i+2^{n-1}} \mbox{ for } i\geq 1 \mbox{, and } 
\Delta=p_{0}-p_{2^{n-1}}\mbox{.}
\end{equation}
Such states \cite{dur00b} play the role of generalized multi-qubit Werner-states
and have the nice property that the positivity
of the partial transposition in respect to a subset $A$  is 
easily checked:
$\rho^{T_{A}}\geq 0 \Leftrightarrow \Delta\leq 2 p_{\mu}$ (where $\vec{\mu}$ 
is the binary array  with $\vec{\mu}_{i}=1$ iff $i\in A$).

For such class of states the $P$-correlation (Eq. \ref{eq:LSGGHZ})
only depends on the parameter $\Delta$ (for odd $n$).
\begin{equation}
L_{P}(\Delta)=2^{n-1}\Delta^{2}+ \delta_{n,\mathrm{even}} 
    \left(2\sum_{\mu=0}^{2^{n-1}-1}(-1)^{|\vec{x}_{i}|} 
    p_{i}+\Delta\right)^{2}
\end{equation}

In the remaining of this section we use the previous results 
to compare the performance of the correlation  strength with
other entanglement criteria:

\begin{itemize}
\item For noisy GHZ:  $p\ketbrad{GHZ}+(1-p) \frac{1}{d}\id$
    $\left\{
    \begin{array}{ll}
        L_{P}(\rho) \rightarrow & 
        p>(2^{n-1}+\delta_{n,\mathrm{even}})^{-\frac{1}{2}}\\
        \mathrm{Bell \;Ineq.} \rightarrow & p>(2^{n-1})^{-\frac{1}{2}} \\
        \mathrm{NPPT} \rightarrow & p> (2^{n-1}+1)^{-1}   \\
    \end{array}
     \right.
     $

  \item $L_{S}(\rho)>1  \nLeftrightarrow 
     \bra{GHZ}\rho\ket{GHZ}>\frac{1}{2}\Rightarrow$ N-distillability.
     That is, the correlation strength criterion is neither stronger nor weaker 
     than the MES-overlap criterion.
     
  \item The correlation strength criterion is neither stronger 
   nor weaker than realignment  criterion \cite{rudolph02,chen03}.  
   It is known that the realignment method can detect some PPT bound 
   entangled states, while the correlation strength can not.
   On the other hand, there are simple examples \textemdash 
   even in the 2x2 case   (see \cite{chen03})\textemdash,
   where the entanglement is not detected by the realignment criteria, 
   but $L_{\{1,2\}}>1$.

   \item  It detects NPPT multi-partite `bound' 
   entanglement\cite{dur01}, $\left\{
    \begin{array}{ll}
        L_{P}(\rho) \rightarrow & 
        N\geq 8\\
        \mathrm{Bell \;Ineq.} \rightarrow & N\geq 8\\
        \mathrm{NPPT} \rightarrow & N\geq  4        
    \end{array}
     \right.
     $

     \item If $\rho$ has $m$ or more positive partial transposes
     $L_{P}(\rho)<L_P(\Delta_{c})$  where 
     $\Delta_{c}=\frac{1}{m+1}$.
     
     \item   k-separability: 
     If a  GHZ-diagonal state can be written as
     $\rho=\sum_{i}p_{i}\rho_{i}^{1}\otimes\ldots\otimes 
     \rho_{i}^{k}$ there will be  $m=2^{k-1}-1$ PPT, and one arrives to 
     $L_{P}(\rho)\leq 2^{n-2k+1}$.
     This coincides with the strongest upper-bound obtained from 
     quadratic Bell-Inequalities \cite{uffink02} in restricted experimental
      settings \cite{nagata02}.
     Using the property
     $L_{S}(\rho_{A}\otimes\rho_{B})=L(\rho_{A})L(\rho_{B})$ it is also 
     possible to derive k-separability criteria for general states
     \cite{calsamiglia04}.

\end{itemize}

\section{Towards a complete set of invariants}
\label{sec:other_invariants}

The local invariants \(L_S\) do not form a complete set of invariants,
\ie they do not contain \emph{all} information about the entanglement
properties of a given state. However, the formalism used in this paper
allows us to identify a larger class of invariants, many of which also
have a straight-forward geometrical interpretation.

From the proof of Theorem~\ref{theorem:one}, it follows that the
transformation properties of the subspaces \(\V_S\) are closely
related. In order to show how this can be used for the construction of
invariants, we first define the $S$-correlation tensor \(C_S\), which
is composed of the components of the projection \(\xi_S\) in
(\ref{eq:decompose_rho}), 
\begin{equation}
  \label{eq:correlation_tensor}
  C_S = \left( %
   \left \langle %
     \prod_{a \in S} \sigma^{(a)}_{i_a} %
   \right\rangle_\rho%
  \right)_{i_a > 0}.
\end{equation}
One can easily see that a contraction of two such tensors
with respect to a index \(i_{\nu}\) at the same position is invariant
under local unitary operations, \ie under orthogonal transformation
\(\mathcal{O}\) which affect this index:
\begin{equation}
  \label{eq:verjuengung}
  \begin{split}
    \sum_{i_\nu} c_{\ldots i_\nu \ldots} c_{\ldots i_\nu \ldots} &=
    \sum_{i_\nu,i''_\nu} \delta_{i_\nu,i''_\nu} c_{\ldots i_\nu
      \ldots} c_{\ldots 
      i''_\nu\ldots} \\%
    & = \sum_{i'_\nu,i_\nu,i''_\nu} \mathcal{O}_{i'_\nu
      i_\nu} c_{\ldots i_\nu \ldots} \mathcal{O}_{i'_\nu i''_\nu} c_{\ldots
      i''_\nu 
      \ldots}\\%
    & = \sum_{i'_\nu} \tilde c_{\ldots i'_\nu \ldots} \tilde c_{\ldots
      i'_\nu \ldots} 
  \end{split}
\end{equation}
Any complete contraction of correlation tensors, \ie a
polynomial in the expansion coefficients, in which indices are either
zero or summed up pairwise, is thus a local invariant. Examples for
such polynomials are \(c_{0jk} c_{ij0} c_{i0k}\),
\(c_{0j00} c_{ij0l} c_{ij'k0} c_{cj'kl}\) (where, as usual, the sum is
taken over all indices which occur twice), the
correlation strengths \(L_S\), and other objects which can be
interpreted as scalar products, such as 
the
scalar product of \(\rho_{a_1a_2a_3}\) with the tensor product of
\(\rho_{\{a_1\}}, \rho_{\{a_2\}}\) and \(\rho_{\{a_3\}}\), 
\begin{equation}
      \label{eq:X_ABC}
      \langle \rho_{\{a_1\}} \otimes \rho_{\{a_2\}} \otimes
      \rho_{\{a_3\}}, \rho_{\{a_1a_2a_3\}}  \rangle
        =  \sum_{i,j,k > 0} c_{i00} c_{0j0} c_{00k} c_{ijk}.
\end{equation}
Unfortunately, it is not possible to construct a complete set
of local invariants using the construction above: already for the case of two
qubits, there are seven independent invariants which can be written as
contraction of correlation tensors; the two remaining invariants are
functions of the determinant and sub-determinants of the correlation
tensor \cite{englert-metwally:99}.


Following the ideas of \cite{grassl-et-al:1998} we can can generalize
the above constructions in order to obtain a complete set of
polynomial invariants in the coordinates $c_{i_1\ldots i_n}$:

Any homogeneous polynomial $f$ of degree $k$, i.e.
\begin{equation} 
  \label{PolynomInvariant}
  f(c_{i_1\ldots
  i_n})= \sum_{{i^1_1,\ldots,i^1_n}\atop{i^2_1,\ldots,i^k_n }} 
  M_{i^1_1,\ldots,i^1_n,i^2_1,\ldots,i^k_n}\;
  c_{i^1_1\ldots i^1_n}  
  c_{i^2_1\ldots i^2_n}\cdots c_{i^k_1\ldots i^k_n},
\end{equation}
is invariant under local unitary transformation,
{\it iff} the corresponding observable 
\begin{equation}
  \label{PolynomObservable}
  M = \sum_{{i^1_1,\ldots,i^1_n}\atop{i^2_1,
    \ldots,i^k_n }} M_{i^1_1,\ldots,i^1_n,i^2_1,\ldots,i^k_n}\;
  \sigma^{(1)}_{i^1_1} \ldots \sigma^{(n)}_{i^1_n} \sigma^{(1)}_{i^2_1}
  \ldots \sigma^{(n)}_{ i^k_n}
\end{equation}
on $\left(\mathcal{H}^{(1)} \otimes \ldots \otimes\mathcal{H}^{(n)}
\right)^{\otimes k}$ commutes with all unitaries of the form
$(U_1^{(1)}\otimes \ldots \otimes U_n^{(n)})^{\otimes k}$. For example, the
proposed invariants $L_S$ are homogeneous polynomials of
degree $2$ for which the corresponding observable is
$M_S=\sum_{\sigma \in \mathcal{B}_S}\, \sigma \otimes \sigma$.  The
map is a vector space isomorphism from the algebra of observables $M$
on $\mathcal{H}^{\otimes k}$, with $[M,(U_1^{(1)}\otimes \ldots
\otimes U_n^{(n)})^{\otimes k}]=0$ for all local unitaries, onto the
the algebra of invariant homogeneous polynomials of degree $k$ in the
coordinates $c_{i_1\ldots i_n}$.  A measurement of the observable $M$
on $k$ copies of the state then evaluates the corresponding polynomial
invariant $M$, i.e.  $f_M(c_{i_1\ldots i_n})= \langle M, \rho^{\otimes
  k} \rangle = \text{tr}\, M \rho^{\otimes k}$. 
  

Note that the orbits of the compact
group of local unitaries can be separated by some polynomial
invariants (see \cite{Onishchik}, p.133).  This means that two states
$\rho_1$ and $\rho_2$ can be transformed into each other by local
unitaries if and only if the evaluations of all polynomial invariants
$f$ coincide.  Moreover, the algebra of invariant polynomials is
finitely generated and  the generators can be chosen to be
homogeneous polynomials.  In order to find a complete set of
invariants, that generates the whole algebra of polynomial invariants,
it is therefore sufficient to find the generators of the algebra of
invariant homogeneous polynomials for different degrees $k$, or
equivalently the algebra of hermitian matrices $M$, that commute with
$(U^{(1)}_1\otimes \ldots \otimes U^{(n)}_n)^{\otimes k}$ for all
local unitaries $U^{(1)}_1\otimes \ldots \otimes U^{(n)}_n$.
The proposed set of invariants $\{L_{S}\}$ are then an example of 
generating set of homogeneous polynomials of degree $k=2$.

In \cite{grassl-et-al:1998} the construction of a set of (possibly
non-hermitean) generators $F$ can be found. From it,  a set of
hermitean generators can be obtained by symmetrization:
$M_1=\frac{F+F^\dagger}{2}$ and $M_2=\frac{F-F^\dagger}{2i}$.  For
each generator $M$ the coefficients
$M_{i^1_1,\ldots,i^1_n,i^2_1,\ldots,i^k_n}$ then can be computed from
Eq.~(\ref{PolynomObservable}), which determines a set of generators
$f_M$ for the invariant ring of degree $k$ by
Eq.~(\ref{PolynomInvariant}).  

\section{Conclusions}
\label{sec:summary}

In this paper, we have investigated local invariants in terms of
correlation operators. The related geometric
interpretation of the space of coherence vectors and its LU-invariant
subspaces allowed us to explicitly write down a subclass of such
invariants.  Using a convexity argument, we have given a simple
 entanglement criterion for multi-partite states.

We have shown that the criterion fails to detect important classes of 
entangled states \textemdash including some NPPT states.
Nevertheless, we think that our criterion is of interest.
The criterion is  simple and with a clear interpretation.
It is easily computable even for large number of qubits 
and some closed forms can be obtained. 
In particular they are more efficiently computable than the 
PPT-criteron that involves a matrix diagonalization. This might be especially usefull
in the study of entanglement in many-body systems.
On the other hand, $L_{P}$ gives information on all partitions so that
one does not need to check the positivity of all $2^{n-1}-1$ possible 
partial transpositions.

From the experimental point of view our criterion also offers some 
advantages. Firstly, the proposed criterion are state-independent: 
it does not require previous knowledge of the state nor
optimization over possible experimental-settings, like in the 
case of Bell-inequalities or general entanglement witnesses.
To measure the $S$-correlation $L_{S}(\rho)$  efficiently 
one can measure a single observable 
$M_S=\sum_{\sigma \in \mathcal{B}_S}\, \sigma \otimes \sigma$
on two copies $\rho\otimes\rho$ requiring only bi-local operations
between equivalent qubits in each of the copies.
Alternatively one can opt for a more straightforward approach
and measure the correlation functions
$ \left\langle \sigma_{i_1}^{(1)} \tensortimes \sigma_{i_2}^{(2)}
    \tensortimes
    \cdots \tensortimes \sigma_{i_n}^{(n)} \right\rangle_\rho$,
as one does in quantum tomography,
with the advantage that entanglement might be detected, \ie $L_{S}$ 
becomes larger than one, long before the 
complete tomography is over.

Finally, it remains an open question whether 
the values achieved by the correlation strengths are related to
the distillability properties of the state.

\section*{Acknowledgements}
This work has been supported by the Deutsche Forschungsgemeinschaft 
(Schwerpunkt QIV)  and the European Union
(IST-2001-38877,-39227).

\end{document}